\documentclass[10pt]{article}

\usepackage{amsmath} %OK
\usepackage{amssymb} %OK
\usepackage{graphicx} %OK
\usepackage{color}  %OK
\usepackage{amsfonts} %OK
\usepackage{ulem}

\usepackage{setspace} % NEED UPLOAD (OK)
\usepackage{cite} % NEED UPLOAD (OK)
\usepackage[mathlines, displaymath, right]{lineno} % NEED UPLOAD (OK)

\date{}
\begin{document}

% Title must be 150 characters or less
\begin{flushleft}
{\Large
%\textbf{Excitatory and Inhibitory Interaction Associated with Non-Uniform Spatial Structure 
%Establishes Energy-Efficient Behavior in Frog Choruses: 
%Theoretical Evidence Using a Simple Mathematical Model}
%}
\textbf{Excitatory and inhibitory interactions affect the balance of 
chorus activity and energy efficiency in the aggregation of male frogs:  
Theoretical study using a hybrid dynamical model}
}
\\
Ikkyu Aihara$^{1, \ast}$, 
Daichi Kominami$^{2}$, 
Yushi Hosokawa$^{2}$, 
Masayuki Murata$^{2}$

\bf{1} Faculty of Engineering, Information and Systems, University of Tsukuba, Ibaraki 305-8573, Japan
\\
\bf{2} Graduate School of Information Science and Technology, Osaka University, Osaka 565-0871, Japan
\\
$\ast$ E-mail: aihara@cs.tsukuba.ac.jp
~
%\\
%Key words: Nonlinear dynamics, Acoustic communication, Satellite behavior, Hybrid dynamical model, Spatial structure
\end{flushleft}

\clearpage

%\begin{keywords}
%Animal Behavior, Synchronization, Phase Oscillator Model, Bayesian Estimation
%\end{keywords}

\section*{Abstract}

We theoretically study the role of excitatory and inhibitory interactions
in the aggregations of male frogs.
In most frogs, males produce sounds to attract conspecific females, 
which activates the calling behavior of other males and results in collective choruses.
While the calling behavior is quite effective for mate attraction, it requires high energy consumption.
In contrast, satellite behavior is an alternative mating strategy 
in which males deliberately stay silent in the vicinity of a calling male 
and attempt to intercept the female attracted to the caller, 
allowing the satellite males to drastically reduce their energy consumption 
while having a chance of mating.
Here we propose a hybrid dynamical model 
in which male frogs autonomously switch among three behavioral states 
(i.e., calling state, resting state, and satellite state)
due to the excitatory and inhibitory interactions.
Numerical simulation of the proposed model demonstrated that 
(1) both collective choruses and satellite behavior can be reproduced 
and (2) the satellite males can prolong the energy depletion time of the whole aggregation 
while they split the maximum chorus activity into two levels over the whole chorusing period.
This study theoretically highlights the trade-off between energy efficiency and chorus activity 
in the aggregations of male frogs driven by the multiple types of interactions.

\begin{flushleft}
{\bf Key words:} Nonlinear dynamics, Acoustic communication, Satellite behavior, Hybrid dynamical model, Spatial structure
\end{flushleft}

\clearpage

\section{Introduction}

Animals aggregate for various purposes such as foraging and mating. 
In the aggregations, energy efficient behavior is observed. 
For instance, an aggregations of ants consists of active and inactive individuals \cite{Bonabeau_1996},
and the switching between the two modes 
likely improves the long-term performance of the whole aggregation \cite{Hasegawa_2016}; 
bats emit ultrasounds to locate surrounding objects by hearing returning echoes \cite{Griffin_1958}.
Indoor experiments demonstrated that flying bats eavesdrop the echos of the preceding individual \cite{Chiu_2008}, 
allowing the follower to reduce energy invested in sound production.
Thus, energy efficient behavior would be common in the aggregations, 
but this phenomenon has not been studied in detail for many species.
Further empirical and theoretical studies are required to analyze 
how energy efficient behaviors of individual animals contribute to the performance of the aggregation as a whole.

\begin{figure*}
  \begin{center}
	\includegraphics[width=1.0\textwidth]{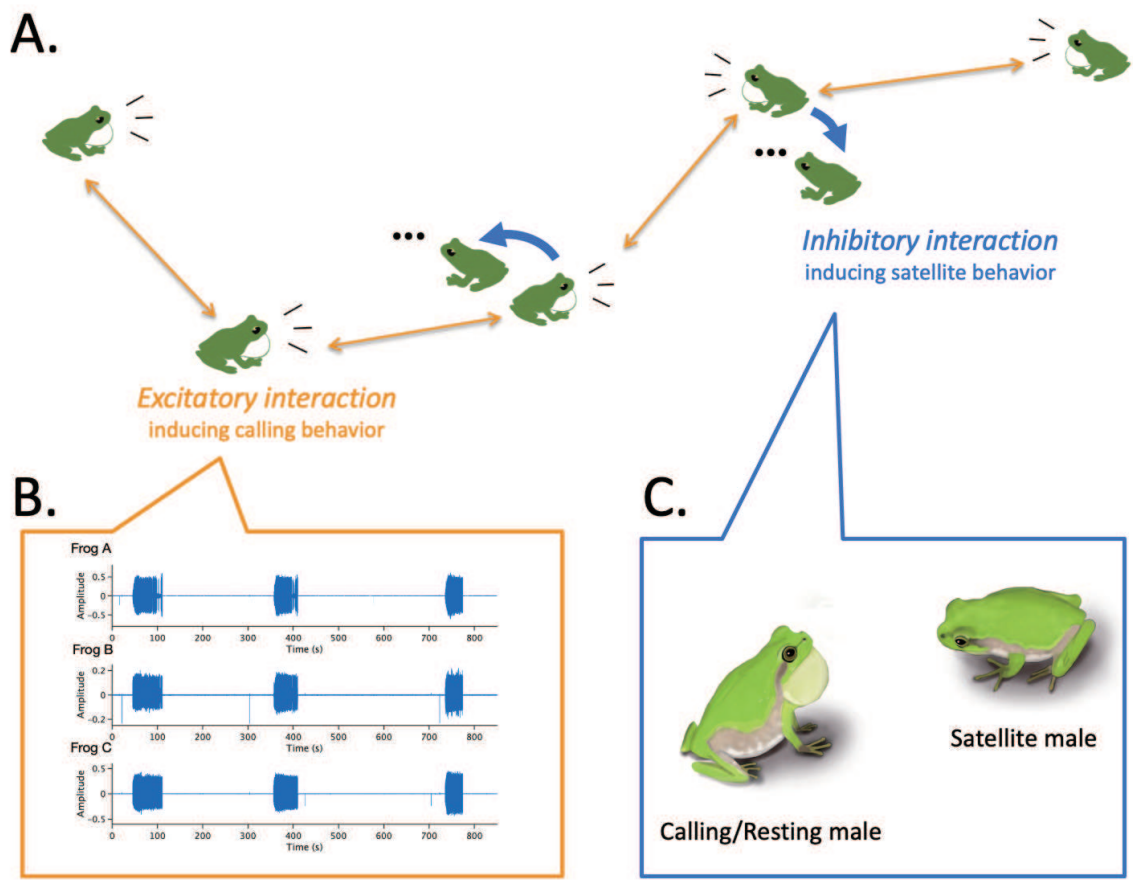}
	\end{center}
	\caption{
	Excitatory and inhibitory interactions in an aggregation of male frogs. 
	(A) Schematic diagram of the aggregation. 
	In this study, we focus on excitatory interaction inducing calling behavior of distant males
	as well as inhibitory interaction inducing satellite behavior of neighboring males. 
	(B) Audio data of male frogs ({\it Hyla japonica}) 
	obtained from our previous studies (Aihara et al., 2011 and 2019). 
	In general the calls of male frogs activate the calling behavior of other males,
	which elicits a collective structure called a unison bout 
	in which males almost synchronize the onset and end of their calling bouts.
	We refer to the interaction as {\it excitatory interaction} among male frogs.
	(C) Satellite behavior in male frogs. 
	As an alternative mating strategy, a male deliberately stays silent in the vicinity of a calling male 
	to intercept a female attracted to the caller. 
	We refer to the relationship as {\it inhibitory interaction} among male frogs.
	}
	\label{fig:concept}
\end{figure*}

In this study, we focus on both energy-consuming and energy-efficient behavior 
in the aggregations of male frogs (Figure \ref{fig:concept}A).
Calling behavior is an energy-consuming behavior in which 
males produce successive sounds by inflating and deflating a large vocal sac 
to attract conspecific females \cite{Gerhardt_2002, Wells_2007}.
Because calling males lose much weight in one night \cite{Gerhardt_2002, MacNally_1981, Cherry_1993, Murphy_1994}, 
this behavior requires males to consume much energy. 
The calls of male frogs generally activate the calling behavior of other males
and elicits a collective structure called as unison bout 
in which males almost synchronize the onset and end of their calling bouts 
\cite{Gerhardt_2002, Wells_2007, Whitney_1975, Schwarts_1994, Jones_2014, Aihara_2019}
(Figure \ref{fig:concept}B).
In this study, we refer to the interaction activating calling behavior of other males as {\it excitatory interaction}.
Contrarily, satellite behavior is an energy-efficient behavior
in which males deliberately stay silent in the vicinity of a calling male (Figure \ref{fig:concept}C)
and attempt to intercept a female attracted to the caller \cite{Gerhardt_2002}.
This behavior allows the males to reduce energy consumption 
because satellite males do not produce any call.
In this study, we refer to the interaction inducing satellite behavior of neighbors as {\it inhibitory interaction}.
Empirical studies showed that calling behavior is dominant in low-density aggregations 
while satellite behavior becomes common in high-density aggregations \cite{Gerhardt_2002}, 
suggesting the dynamical choice of behavioral types depending on surrounding conditions.
This study aims to theoretically examine how the coexistence of excitatory and inhibitory interactions 
contributes to chorus activity and energy efficiency in the aggregations of male frogs.

% INSERT FIGURE 

\section{Mathematical Modeling}

We propose a mathematical model incorporating both excitatory and inhibitory acoustic interactions among male frogs
as an extension of our study \cite{Aihara_2019}. 
Specifically, we model a calling state, a resting state, and a satellite state as separate deterministic models, 
and then formulate the transitions among the three states as stochastic processes (Figure \ref{fig:framework}). 
A calling state describes the behavior in which a male frog produces successive sounds 
to attract conspecific females. 
The production of sounds requires the vigorous inflation and deflation of a large vocal sac at a high repetition rate
\cite{Gerhardt_2002, Wells_2007}, 
and calling males lose much weight in one night \cite{Gerhardt_2002, MacNally_1981, Cherry_1993, Murphy_1994}.
Based on these features, we assume that (1) the energy of a calling male decreases  
while his physical fatigue increases, 
and  that (2) the male stops calling when he gets tired. 
A resting state describes the behavior in which a male frog stays silent 
without producing any call during an interval between calling states 
\cite{Gerhardt_2002, Wells_2007, Whitney_1975, Schwarts_1994, Jones_2014, Aihara_2019} 
(Figure \ref{fig:concept}B).
Because of the lower activity, we assume that 
(1) the energy of a resting male remains constant 
while his physical fatigue decreases, 
and that (2-1) the male starts calling when he has rested for enough time and is activated by the calls of other males 
or (2-2) the male frog transits to the satellite state 
when the attractiveness of his neighbor is superior to his own.  
A satellite state describes the behavior in which a male frog stays silent in the vicinity of a calling male
to intercept a female attracted to the caller \cite{Gerhardt_2002}.
Because of the lower activity and mating strategy, 
we assume that 
(1) the energy of a satellite male remains constant 
while his physical fatigue decreases, and 
(2) he transits to a resting state and starts calling again 
when the attractiveness of his neighbor is inferior to his own.

This section is organized as follows.
In Sec. \ref{sec:Definition of Behavioral States},
we introduce the deterministic models of the calling state, the resting state and the satellite state, respectively.
In Sec. \ref{sec:Definition of Transitions}, 
we formulate the transitions among the three states.
In Sec. \ref{sec:Parameter values}, 
we fix the parameters of the proposed model 
based on the behavioral features of male frogs.

\subsection{Formulation of Behavioral States}
\label{sec:Definition of Behavioral States}

We formulate three behavioral states (the calling state, the resting state and the satellite state) 
as separate deterministic models. 
It should be noted that the models of the calling state and the resting state were already proposed 
in our previous study \cite{Aihara_2019}.
Here we newly introduce the model of  the satellite state
based on the model of the resting state.

\begin{figure*}
  \begin{center}
	\includegraphics[width=1.0\textwidth]{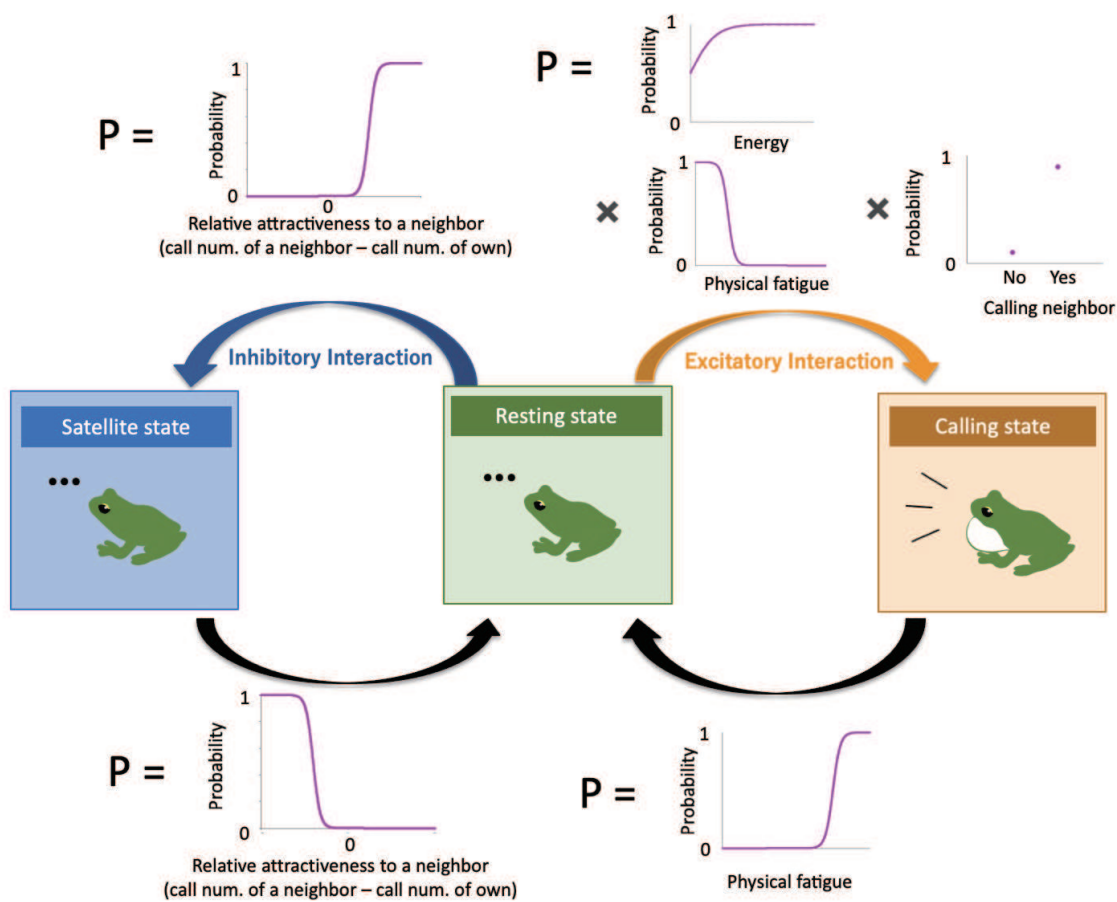}
	\end{center}
	\caption{
	A mathematical model incorporating excitatory and inhibitory interactions among male frogs. 
	Each frog is assumed to be in a calling state, a resting state, or a satellite state, and transit among the three states.
	The transition between the calling state and the resting state is described by a framework proposed by Aihara et al., 2019:
	the resting males with low physical fatigue and high energy have a high probability to be activated by the calls of neighboring males 
	and start calling (corresponding to {\it excitatory interaction})
	while calling males attempt to continue calling until physical fatigue exceeds a threshold value. 
	We newly formulate the transition between the satellite state and the resting state.
	Specifically, male frogs are assumed to transit from the resting state to the satellite state at high probability
	when the attractiveness of their calls is moderately inferior to that of a neighboring male
	(corresponding to {\it inhibitory interaction}). 
	Inversely, male frogs transit from the satellite state to the resting state 
	when the attractiveness of their calls is moderately superior to that of a neighboring male. 
	}
	\label{fig:framework}
\end{figure*}

First, we describe the calling state of the $n$th frog as follows \cite{Aihara_2019}:
\begin{eqnarray}
\frac{d\theta_{n}}{dt} &=& \omega_{n} + \sum_{m \mathrm{~for~} r_{nm} < r_{\mathrm{acoustic}}}^{N} \delta(\theta_{m}) \Gamma_{nm}(\theta_{n}-\theta_{m}),
\label{eq:call_theta}\\
\frac{dT_{n}}{dt} &=& \delta(\theta_{n}),
\label{eq:call_T}\\
\frac{dE_{n}}{dt} &=& -\delta(\theta_{n}),
\label{eq:call_E}
\end{eqnarray}
where 
\begin{eqnarray}
\Gamma_{nm}(\theta_{n}-\theta_{m}) &=& K_{nm} \big[ \sin(\theta_{n}-\theta_{m}) - k \sin(2(\theta_{n}-\theta_{m})) \big].
\label{eq:interaction_term}
\end{eqnarray}
Equation (\ref{eq:call_theta}) is based on a mathematical model called a phase oscillator model 
that is derived from simple assumptions about periodicity  and interaction \cite{Kuramoto_1984}.
The phase oscillator model can qualitatively reproduce various synchronization phenomena in biological systems \cite{Strogatz, Nenkin, BZ, Walk} 
including frog choruses \cite{Aihara_2011, Aihara_2014, Aihara_2019, Aihara_2020}.
We utilize the model to describe periodic calling behavior of male frogs
and also their acoustic interaction.
In Equation (\ref{eq:call_theta}), $\theta_{n}$ is a variable ranging from $0$ to $2\pi$ \cite{Kuramoto_1984} 
and represents the phase of calls produced by the $n$th frog \cite{Aihara_2011, Aihara_2014, Aihara_2019}. 
Specifically, the $n$th frog is assumed to produce a call when the phase $\theta_{n}$ hits $0$.
$\omega_{n}$ is a positive parameter that determines an intrinsic inter-call interval of this frog 
\cite{Aihara_2011, Aihara_2014, Aihara_2019}.
In the second term in the right-hand side of Equation (\ref{eq:call_theta}), 
$\delta(\theta_{m})$ is a delta function satisfying the conditions $\delta(\theta_{m})=\infty$ at $\theta_{m}=0$, 
$\delta(\theta_{m})=0$ otherwise, and  $\int_{t_{m, i}-\epsilon}^{t_{m, i}+\epsilon} \delta(\theta_{m}(t)) dt = 1$ 
($i$ represents the index of the calls produced by the $m$th frog, 
and $\epsilon$ is a positive parameter that is much smaller than the inter-call interval) 
\cite{Aihara_2019}. 
$\Gamma_{nm}(\theta_{n}-\theta_{m})$ is a $2\pi$-periodic function of the phase difference $\theta_{n}-\theta_{m}$
\cite{Kuramoto_1984}.
In addition, $\Gamma_{nm}(\theta_{n}-\theta_{m})$ is given by Equation (\ref{eq:interaction_term})
with two kinds of coupling strength $K_{nm}$ and $k$
because this function can qualitatively reproduce alternating chorus patterns of male frogs %over a short time scale 
\cite{Aihara_2011, Aihara_2019} 
that are generally observed within each chorusing bout
\cite{Gerhardt_2002, Wells_2007, Brush_1989, Jones_2014}.
$r_{nm}$ is the distance between the $n$th and $m$th frogs, 
and $r_{\mathrm{acoustic}}$ is a threshold within which male frogs can acoustically interact with each other.
Specifically, we assume that the $n$th frog pays attention to the first and second nearest callers 
based on the previous studies reporting that male frogs acoustically interact with a few neighbors in an aggregation 
\cite{Gerhardt_2002, Wells_2007, Brush_1989, Aihara_2016, Jones_2014}.
Consequently, the second term in the righ-hand side of Equation (\ref{eq:call_theta})
describes an instantaneous selective interaction between the $n$th and $m$th frogs 
(In other words, when the $m$th frog produces a call at $\theta_{m}=0$, 
the phase of the $n$th frog that selectively pays its attention to the $m$th frog is instantaneously affected by the call).
In Equations (\ref{eq:call_T}) and (\ref{eq:call_E}), $T_{n}$ and $E_{n}$ describe
the physical fatigue and energy of the $n$th frog
and are restricted to the ranges of $0 \le T_{n} \le T_{\mathrm{max}}$ and $0 \le E_{n} \le E_{\mathrm{max}}$, 
respectively \cite{Aihara_2019}. 
Because of the definition of $\delta(\theta_{n})$, 
the physical fatigue $T_{n}$ is incremented by $1$ 
and the energy $E_{n}$ is decremented by $1$ 
when the $n$th frog produces a call at $\theta_{n}=0$ \cite{Aihara_2019}.
As explained at the beginning of this section,
our model assumes that males in the calling state can transit to the resting state
(Figure \ref{fig:framework}).

Next, we describe the resting state and the satellite state as follows \cite{Aihara_2019}:
\begin{eqnarray}
\frac{d\theta_{n}}{dt} &=& 0,
\label{eq:sleep_theta}\\
\frac{dT_{n}}{dt} &=& -\alpha.
\label{eq:sleep_T}\\
\frac{dE_{n}}{dt} &=& 0,
\label{eq:sleep_E}
\end{eqnarray}
Here we utilize the same framework
because a male frog is assumed to stay silent during both states.
Equation (\ref{eq:sleep_theta}) means that the phase $\theta_{n}$ remains constant without hitting $0$ 
and therefore a model frog does not produce any call. 
Equations (\ref{eq:sleep_T}) and (\ref{eq:sleep_E}) 
describe the time evolution of the physical fatigue $T_{n}$ and the energy $E_n$, respectively.
These equations mean that the physical fatigue $T_{n}$ decreases 
and the energy $E_n$ remains constant, 
which is consistent with the assumed behavioral features of both states.
As explained at the beginning of this section,  
model frogs in the resting state can transit to the calling state or the satellite state 
while the frogs in the satellite state can transit only to the resting state
(Figure \ref{fig:framework}).

% INSERT FIGURE 

\subsection{Formulation of Transitions}
\label{sec:Definition of Transitions}

Here we formulate the transitions among three behavioral states 
(i.e., the calling state, the resting state and the satellite state).
First, we utilize an integer $s_{n}$ to discriminate among the three states:
namely, the calling state of the $n$th frog is described as $s_{n}=1$, 
the resting state is described as $s_{n}=0$, 
and the satellite state is described as $s_{n}=-1$.
Second, we propose a stochastic mathematical model in which each male spontaneously switches its state
depending on his current condition and also the interaction with other males.

The transition between the calling state and the resting state is formulated
according to our previous study \cite{Aihara_2019}. 
Specifically, our previous model \cite{Aihara_2019} assumes that 
(1) a male frog in the calling state has a high probability to continue calling when his physical fatigue is low, 
and 
(2) a male frog in the resting state has a high probability to be activated by the calls of neighboring males when his energy is high and his physical fatigue is low. 
Based on the first assumption, the probability of the transition from the calling state to the resting state 
is given as follows \cite{Aihara_2019}: 
\begin{eqnarray}
P^{\mathrm{call \to rest}}_{n} &=& G_{1}(T_{n}), 
\label{eq:p_call-rest}
\end{eqnarray}
where
\begin{eqnarray}
G_{1}(T_{n}) &=& \frac{1}{\exp(-\beta_{\mathrm{fatigue}} (T_{n} - \Delta T))+1}.
\label{eq:G_1}
\end{eqnarray}
$G_{1}(T_{n})$ is a logistic function that increases depending on a parameter $\beta_{\mathrm{fatigue}}$.
Another parameter $\Delta T$ represents the inflection point of the logistic function.
Equations (\ref{eq:p_call-rest}) and (\ref{eq:G_1}) mean 
that the probability to stop calling becomes much higher 
when the physical fatigue $T_{n}$ exceeds the parameter $\Delta T$ (Figure \ref{fig:framework}).
Next, based on the second assumption, 
the probability of the transition from the resting state to the calling state is given as follows \cite{Aihara_2019}: 
\begin{eqnarray}
P^{\mathrm{rest \to call}}_{n} &=& G_{2}(T_{n}) G_{3}(E_{n}) G_{4}(\vec{s}_{n}^{\mathrm{neighbor}}),
\label{eq:p_rest-call}
\end{eqnarray}
where
\begin{eqnarray}
G_{2}(T_{n}) &=& \frac{1}{\exp(\beta_{\mathrm{fatigue}} (T_{n} - (T_{\mathrm{max}} - \Delta T)))+1},
\label{eq:G_2}\\
G_{3}(E_{n}) &=& -\frac{2}{\exp(\beta_{\mathrm{energy}} E_{n})+1} + 1,
\label{eq:F}
\end{eqnarray}
\begin{flalign}
G_{4}(\vec{s}_{n}^{\mathrm{neighbor}}) &= 
  \begin{cases}
    p_{\mathrm{high}}  & \text{(If a vector $\vec{s}_{n}^{\mathrm{neighbor}}$ has one or more elements of $1$)},\\
    p_{\mathrm{low}} & \text{(If a vector $\vec{s}_{n}^{\mathrm{neighbor}}$ has no element of $1$)}.
  \end{cases}
\label{eq:H}
\end{flalign}
$G_{2}(T_{n})$ and $G_{3}(E_{n})$ represent the effects of the physical fatigue $T_{n}$ and the energy $E_{n}$ on the transition.
Specifically, $G_{2}(T_{n})$ is a logistic function that increases depending on a parameter $\beta_{\mathrm{fatigue}}$
when the physical fatigue $T_{n}$ decreases, 
and $G_{3}(E_{n})$ is another logistic function that decreases depending on a parameter $\beta_{\mathrm{energy}}$
when the energy $E_{n}$ decreases (Figure \ref{fig:framework}).
In Equation (\ref{eq:H}), $\vec{s}_{n}^{\mathrm{neighbor}}$ represents the states of males
that are closer to the focal frog than a threshold value of $r_{\mathrm{acoustic}}$; 
$p_{\mathrm{high}}$ and $p_{\mathrm{low}}$ are positive parameters 
that satisfy the relationship $p_{\mathrm{high}} \gg p_{\mathrm{low}} > 0$.
Subsequently, $G_{4}(\vec{s}_{n}^{\mathrm{neighbor}})$ is a discrete function 
that takes $p_{\mathrm{high}}$ when one or more males are calling in his interaction range %$r_{\mathrm{acoustic}}$
(or it takes $p_{\mathrm{low}}$ when no males are calling). 
Equations (\ref{eq:p_rest-call})--(\ref{eq:H}) assume that 
male frogs with lower physical fatigue and higher energy 
have a high probability to be activated by the calls of other males 
and then transit to the calling state, 
corresponding to {\it excitatory interaction} that is a focus of this study.

Next, we formulate the transition between the resting state and the satellite state 
as another stochastic process 
by focusing on relative attractiveness between neighboring males. 
In frog choruses, temporal traits of calls vary among individual males 
and work as an important indicator to determine their attractiveness towards conspecific females 
\cite{Gerhardt_2002, Wells_2007}. 
In particular, the number of calls dominantly affects the attractiveness.
Playback experiments using various frog species showed 
that loudspeaker broadcasting more calls per unit time 
have a higher probability of attracting females
than the loudspeaker broadcasting less calls \cite{Ryan_1992}.
Based on this feature, we treat the number of calls as a representative indicator of the attractiveness  
and formulate the probability of the transition from the resting state to the satellite state as follows: 
\begin{eqnarray}
P^{\mathrm{rest \to satellite}}_{n} &=& F_{1}(N_{k}-N_{n}), 
\label{eq:p_rest-satellite}
\end{eqnarray}
where
\begin{eqnarray}
F_{1}(N_{k}-N_{n}) &=& \frac{1}{\exp(-\beta_{\mathrm{satellite}}(N_{k}-N_{n} - \Delta N))+1}.
\label{eq:F_1}
\end{eqnarray}
Empirical studies demonstrated that 
males intermittently start and stop calling in various frog species
(Figure \ref{fig:concept}(B) for the case of {\it Hyla japonica})
\cite{Gerhardt_2002, Wells_2007, Whitney_1975, Schwarts_1994, Jones_2014, Aihara_2019}.
To precisely capture this temporal feature as well as take the effect of the call number into consideration, 
we describe the number of calls included in the adjacent calling bout as an integer $N_{n}$
and utilize it as the current attractiveness of the $n$th frog.
$F_{1}(N_{k}-N_{n})$ is a logistic function with a steepness $\beta_{\mathrm{satellite}}$
that takes larger value 
when the difference of the call number ($N_{k}-N_{n}$) is larger than a threshold value $\Delta N$. 
Here we choose a specific male as the $k$th frog in Equations (\ref{eq:p_rest-satellite}) and (\ref{eq:F_1}) 
if he positions within a threshold of $r_{\mathrm{satellite}}$
and engages in the calling state or the resting state.
These assumptions mean that the focal model frog (the $n$th frog) is a satellite of a neighboring male (the $k$th frog) 
and attempts to intercept a female attracted to the neighbor. 
Subsequently, a model frog switches from the resting state to the satellite state 
when his attractiveness is inferior to that of the neighboring male, 
corresponding to {\it inhibitory interaction} that is a focus of this study.
Next, we formulate the probability of the transition from the satellite state to the resting state as follows:
\begin{eqnarray}
P^{\mathrm{satellite \to rest}}_{n} &=& F_{2}(N_{k}-N_{n}), 
\label{eq:p_satellite-rest}
\end{eqnarray}
where
\begin{eqnarray}
F_{2}(N_{k}-N_{n}) &=& \frac{1}{\exp(\beta_{\mathrm{satellite}}(N_{k}-N_{n} + \Delta N))+1}.
\label{eq:F_2}
\end{eqnarray}
$F_{2}(N_{k}-N_{n})$ is a logistic function with the steepness $\beta_{\mathrm{satellite}}$ 
and becomes larger when the difference of the call number ($N_{k}-N_{n}$) is lower than the threshold $\Delta N$. 
Equations (\ref{eq:p_satellite-rest}) and (\ref{eq:F_2}) assume that 
a model frog switches from the satellite state to the resting state 
when his attractiveness is superior to that of the neighbor.

There are two choices for the transition from the resting state 
(i.e., the transition from the resting state to the calling state or the satellite state; see also Figure \ref{fig:framework}). 
When determining to which state each model frog transits,
we compare the probability of both transitions 
(i.e., $P^{\mathrm{rest \to call}}_{n}$ and $P^{\mathrm{rest \to satellite}}_{n}$) 
and choose the transition 
with higher probability of happening.

\subsection{Parameter values}
\label{sec:Parameter values}

As explained in Sec. \ref{sec:Definition of Behavioral States} and \ref{sec:Definition of Transitions}, 
we previously proposed a mathematical model on the transition between the calling state and the resting state 
and succeeded in reproducing the occurrence of the collective choruses \cite{Aihara_2019}.
Because this study also focuses on the collective choruses, 
we utilize the same parameter values \cite{Aihara_2019} for the corresponding parts of the proposed model
(i.e., the deterministic model of the calling state 
(Equations (\ref{eq:call_theta})--(\ref{eq:call_E})), 
the deterministic model of the resting state 
(Equations (\ref{eq:sleep_theta})--(\ref{eq:sleep_E})), 
and  the stochastic model of the transition between the calling state and the resting state 
(Equations (\ref{eq:p_call-rest})--(\ref{eq:H}))).

Next, we fix or vary the parameters included 
in the stochastic model of the transition between the satellite state and the resting state
(Equations (\ref{eq:p_rest-satellite})--(\ref{eq:F_2})). 
Because male frogs in the satellite state attempt to intercept a conspecific female attracted to their calling neighbor, 
he needs to stay in the vicinity of the neighbor. 
In our model, this feature is modeled by the parameter $r_{\mathrm{satellite}}$ 
within which a model frog transits to the satellite state (Sec.\ref{sec:Definition of Transitions}). 
Therefore, we fix $r_{\mathrm{satellite}}$ at a small value of $0.4$ m
that allows the satellite male to immediately catch a female attracted to the neighbor. 
Then, we vary the remaining parameters of $\beta_{\mathrm{satellite}}$ and $\Delta N$ 
(see Equations (\ref{eq:p_rest-satellite})--(\ref{eq:F_2}))
that  are related to the transition between the satellite state and the resting state.

The values and meanings of all the parameters are summarized 
in Tables S1 -- S3  in Supplementary Information.

\section{Numerical simulation}
\label{sec:Numerical simulation}

We performed numerical simulations of the proposed model
to analyze how the coexistence of excitatory and inhibitory interactions affects 
chorus activity and energy efficiency in aggregations of male frogs. 
First, to simply test the validity of the proposed model, 
we simulated an aggregation of three model frogs that is the minimum unit 
exhibiting both the collective chorus and satellite behavior 
(Sec. \ref{sec:Small chorus}).
Second, we simulated the aggregations of $10$--$20$ model frogs
that more accurately imitate the aggregations of actual male frogs (Sec. \ref{sec:Larger chorus}).

\subsection{Small aggregation of male frogs}
\label{sec:Small chorus}

\begin{figure*}
  \begin{center}
	\includegraphics[width=1.0\textwidth]{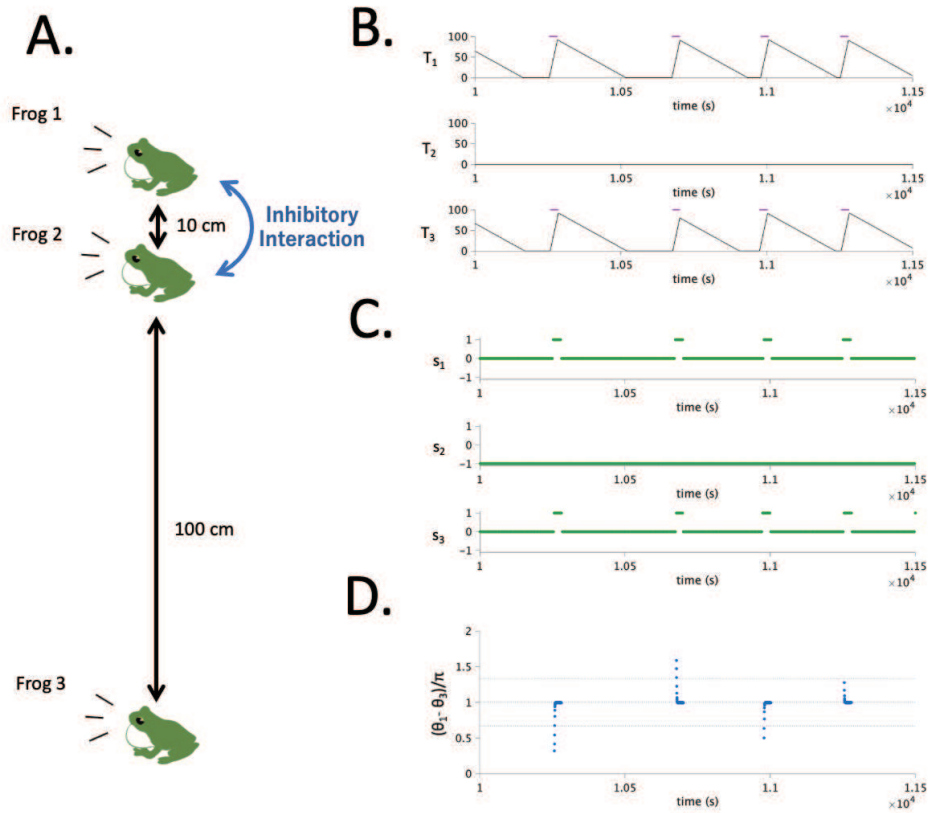}
	\end{center}
	\caption{
	Numerical simulation on the small aggregation of three male frogs. 
	(A) Spatial structure of the frogs. 
	Two of the three frogs are assumed to be positioned in a close distance within which satellite behavior can be induced.
	(B) Time evolution of the physical fatigue $T_{n}$. 
	(C) Time evolution of the state variable $s_{n}$. 
	Calling state, resting state and satellite state are described by $s_{n}=1$, $0$ and $-1$, respectively.
	(D) Time evolution of the phase difference ($\theta_{1}-\theta_{3}$) in each chorus. 
	The phase $\theta_{n}$ describes the call timing of the $n$th frog.
	While Frog 2 engages in the satellite state, Frogs 1 and 3 exhibit collective choruses 
	within which they call alternately in anti-phase. 
	}
	\label{fig:Small_aggregation}
\end{figure*}

For the numerical simulation of a small aggregation, 
we assume that three model frogs are distributed at different inter-frog distances 
along a line (Figure \ref{fig:Small_aggregation}A).
Frogs 1 and 2 are positioned at a close distance of $0.1$ m apart, 
and the pair of Frogs 2 and 3 are positioned at a long distance of $1.0$ m apart.
Because the parameter $r_{\mathrm{satellite}}$ was set at $0.4$ m 
(see Sec. \ref{sec:Parameter values}), 
the inhibitory interaction can occur only in the closest pair (Frogs 1 and 2)
and the excitatory interaction can occur in the other pairs (Frogs 1 and 3 and Frogs 2 and 3).

Numerical simulation of the proposed model demonstrated the occurrence of collective choruses and satellite behavior.
Figure \ref{fig:Small_aggregation} B and C shows the time evolution of the physical fatigue $T_{n}$ and the state variable $s_{n}$ 
with the parameters of $\Delta N = 15$ and $\beta_{\mathrm{satellite}}=0.5$, 
indicating that Frogs 1 and 3 exhibit collective choruses by synchronously switching 
between the resting state ($s_{n}=0$) and the calling state ($s_{n}=1$)
while Frog 2 engages in the satellite state ($s_{n}=-1$) . 
To examine the temporal structure within each chorus, we calculated the phase difference between Frogs 1 and 3. 
Figure \ref{fig:Small_aggregation}  C shows that the phase difference ($\theta_{1}-\theta_{3}$) converged to $\pi$ in each chorus, 
which corresponds to anti-phase synchronization of the two frogs. 
Note that the anti-phase synchronization within choruses 
can be generally observed in frog choruses \cite{Gerhardt_2002, Wells_2007, Brush_1989, Jones_2014}
and likely prevents the males from interfering with each other's calls 
\cite{Schwartz_1987, Bee, Henry_2019}.
Thus, the proposed model can reproduce three behaviors of actual male frogs,
i.e., satellite behavior, collective choruses and also anti-phase synchronization within choruses.

% INSERT FIGURE 
%%

\begin{figure*}
  \begin{center}
	\includegraphics[width=1.0\textwidth]{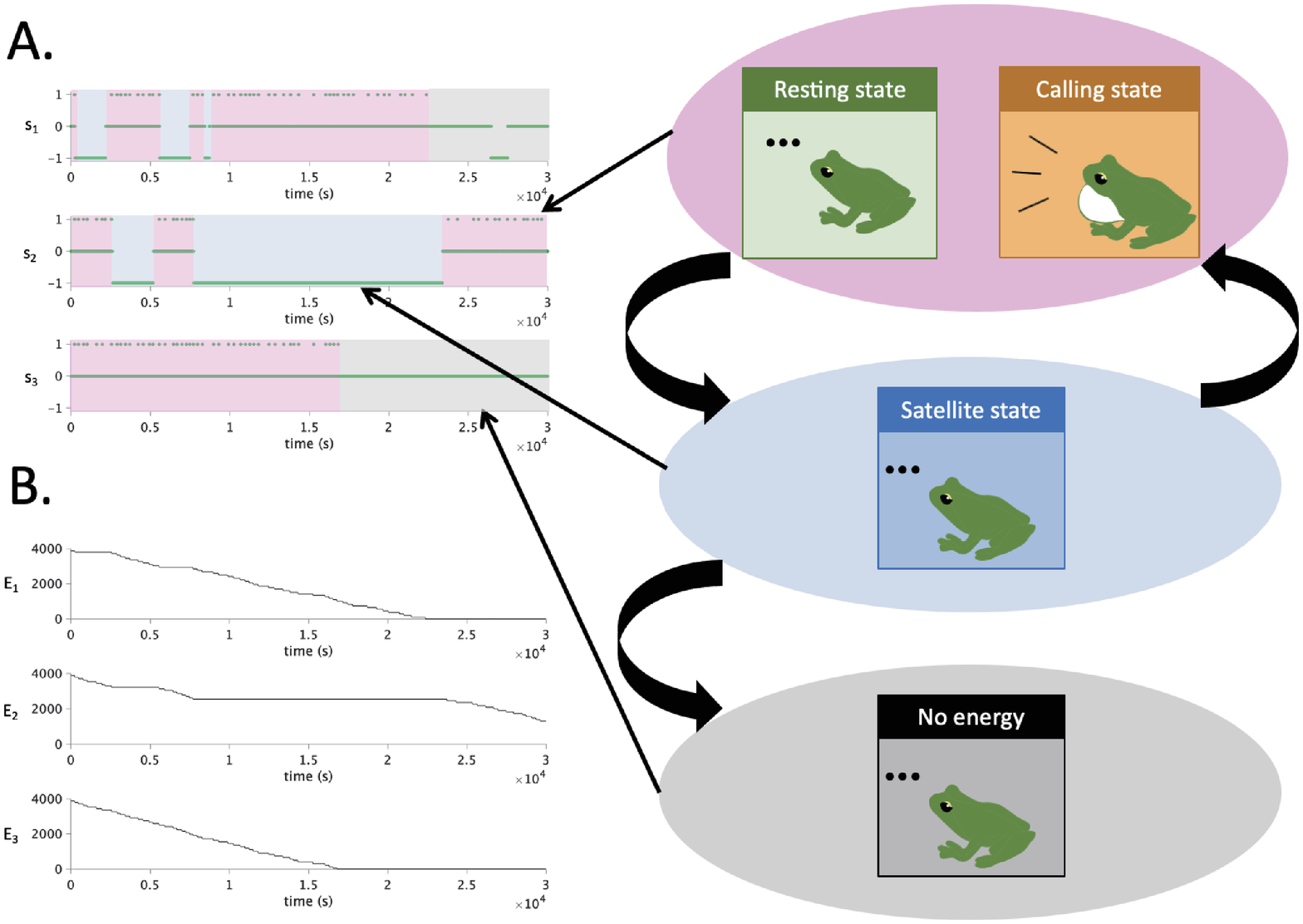}
	\end{center}
	\caption{
	Numerical simulation on the small aggregation over a longer time scale.
	(A) Time evolution of the state variable $s_{n}$. 
	While one of the closest pair (Frog 1 or Frog 2) engages in satellite behavior (light blue region),
	the remaining males join in collective choruses (pink region).
	(B) Time evolution of the energy $E_{n}$.
	The pace of energy consumption is slowed down in the closest pair
	because of intermittent transitions into the satellite state.
	}
	\label{fig:SmallAggre_LongSim}
\end{figure*}

\begin{figure*}
  \begin{center}
	\includegraphics[width=1.0\textwidth]{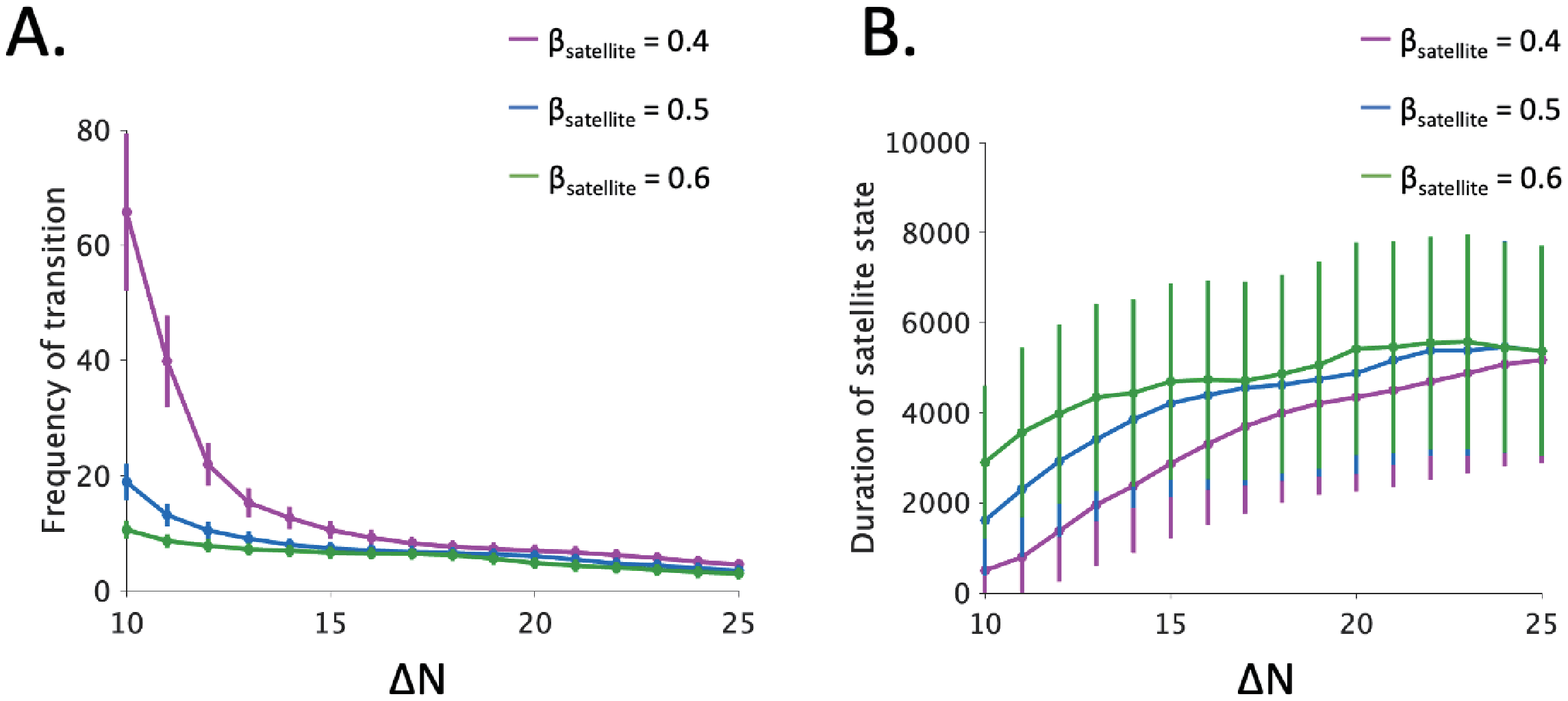}
	\end{center}
	\caption{
	Numerical simulation on the parameter dependency of satellite behavior.
	Parameters $\Delta N$ and $\beta_{\mathrm{satellite}}$ 
	dominantly affect the probability of the transition between the satellite state and the resting state 
	(see Equations (\ref{eq:p_rest-satellite})--(\ref{eq:F_2})).
	(A) The frequency of the transition into the satellite state.
	(B) The duration of the satellite state.
	We repeated the simulation $1000$ times at each parameter set with randomized initial conditions.
	Plots and bars represent the mean and standard deviation of each quantity.
	While the frequency of the transition decreases as the parameter $\Delta N$ increases,
	the duration of the satellite state increases as the parameter $\Delta N$ increases.
	}
	\label{fig:SmallAggre_ParaDep}
\end{figure*}

Next, we analyzed the detailed trait of mode switching over a long time scale. 
Figure \ref{fig:SmallAggre_LongSim}A and B shows the time evolution of the state variable $s_{n}$ and that of the energy $E_{n}$, respectively.
The model frogs can switch between the collective choruses (pink region) and the satellite behavior (light blue region).
Specifically, the satellite male spontaneously switches between the closest pair of Frog 1 and Frog 2 (Figure \ref{fig:SmallAggre_LongSim}A)
and the pace of the energy consumption is slowed down (Figure \ref{fig:SmallAggre_LongSim}B).
To further examine the generality of these results, 
we varied the parameters $\Delta N$ and $\beta_{\mathrm{satellite}}$ that dominantly affect 
the probability of the transition to the satellite state 
(see Equations (\ref{eq:p_rest-satellite})--(\ref{eq:F_2})), 
and repeated the simulation $1000$ times at each parameter value with different initial conditions.
Figure \ref{fig:SmallAggre_ParaDep} shows that 
(1) the frequency of the transition decreases as the parameter $\Delta N$ increases 
and (2) the duration of the satellite state increases as the parameter $\Delta N$ increases. 
Thus, the simulation of the proposed model indicates the dynamical switching 
between the collective choruses and satellite behavior 
while  the temporal feature varies especially depending on the parameter $\Delta N$.

% INSERT FIGURE 
%%
%%

% INSERT FIGURE 
%%

\subsection{Large aggregations of male frogs}
\label{sec:Larger chorus}

\begin{figure*}
  \begin{center}
	\includegraphics[width=1.0\textwidth]{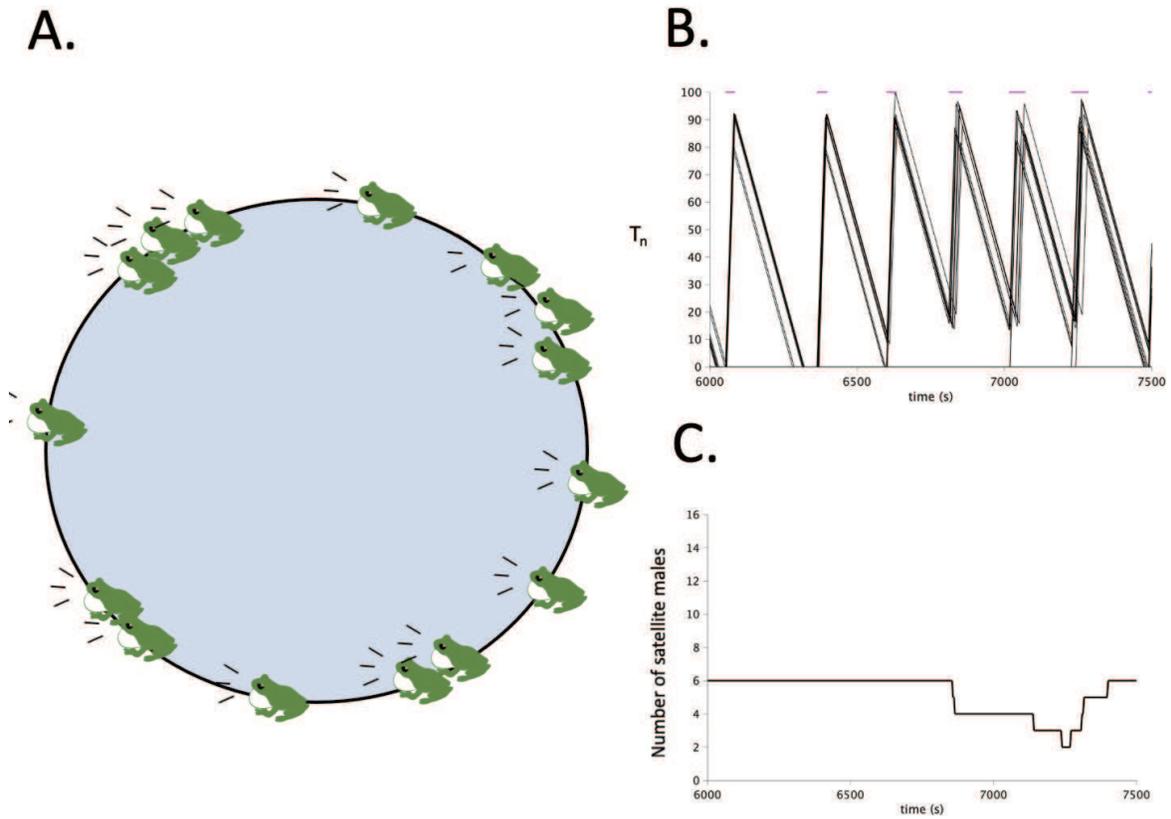}
	\end{center}
	\caption{
	Numerical simulation on a large aggregation of 15 male frogs. 
	(A) Schematic diagram of the positions of male frogs. 
	In the simulation we assume that model frogs are distributed along the edge of a breeding water body, 
	which is a common spatial distribution of male frogs in the wild.
	(B) Time evolution of the physical fatigue $T_{n}$.
	(C) Time evolution of the number of satellite males. 
	While the majority of the model frogs exhibit collective choruses
	by almost synchronizing the dynamics of the physical fatigue $T_{n}$, 
	the remaining frogs engage in satellite behavior. 
	}
	\label{fig:LargeAggre_concept}
\end{figure*}

For the numerical simulation of large aggregations, 
we assume that $10$--$20$ frogs are distributed around a breeding site 
(Figure \ref{fig:LargeAggre_concept}A).
In the wild the positions of male frogs change every night, 
but it is common that they are distributed along the edge of a water body
\cite{Aihara_2014, Aihara_2016, Aihara_2021, Bando_2016, Aihara_2017}.
To simply imitate the spatial distribution as well as the variance among nights, 
we positioned the model frogs along the edge of a circular water body 
and then randomized their positions in each trial of the simulation.

Numerical simulation demonstrated the occurrence of collective choruses and satellite behavior in a large aggregation (Figure \ref{fig:LargeAggre_concept}A).
Figure \ref{fig:LargeAggre_concept}B and C shows the time evolution of the physical fatigue $T_{n}$ 
and that of the number of satellite males, respectively. 
While the majority of the model frogs exhibit collective choruses
by almost synchronizing the dynamics of the physical fatigue $T_{n}$, 
the remaining frogs engage in the satellite behavior.

% INSERT FIGURE
%%

\begin{figure*}
  \begin{center}
	\includegraphics[width=1.0\textwidth]{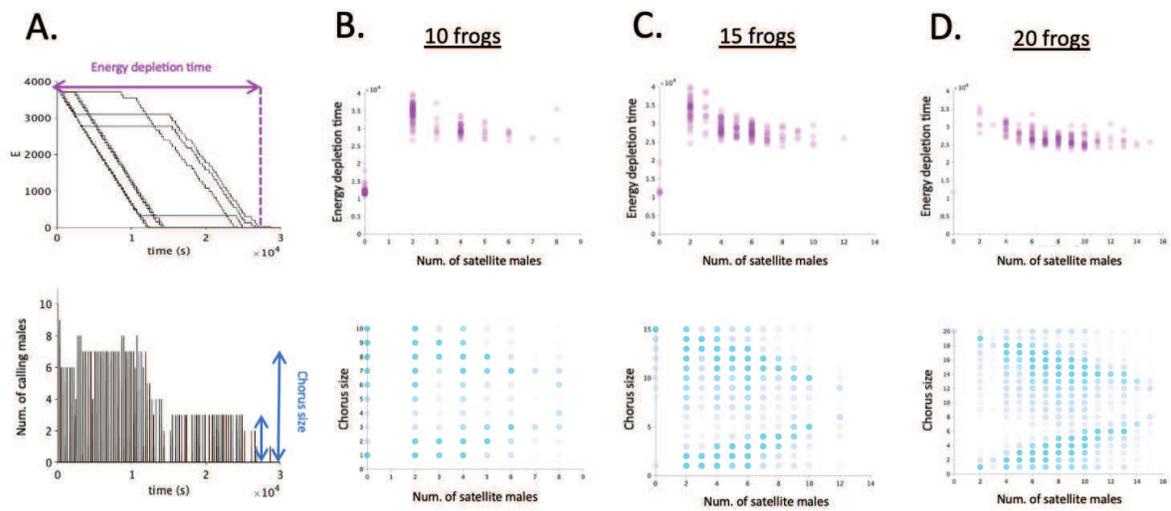}
	\end{center}
	\caption{
	Numerical simulation on the performance of large aggregations.
	(A) Schematic diagram of {\it energy depletion time} and {\it chorus size}.
	(B)--(D) The dependency of {\it energy depletion time} and {\it chorus size}  
	on the number of satellite males 
	in the aggregations of $10$ frogs, $15$ frogs and $20$ frogs, respectively. 
	We repeated the simulation $200$ times in each aggregation with randomized initial conditions, 
	and overlaid translucent plots in each graph.
	While the energy depletion time is drastically prolonged 
	in the aggregations with satellite males (the top panels), 
	the effective chorus size can take two peaks depending on the number of satellite males (the bottom panels).
	}
	\label{fig:LargeAggre_Quality}
\end{figure*}

Next, we numerically evaluated the performance of large aggregations.
Figure \ref{fig:LargeAggre_Quality}A shows the entire time evolution of the energy $E_{n}$
and that of the number of calling males.
While the pace of energy consumption is slowed down due to intermittent transitions into the satellite state 
(the top panel of Figure \ref{fig:LargeAggre_Quality}A), 
several individuals consistently join in collective choruses
(the bottom panel of Figure \ref{fig:LargeAggre_Quality}A).
Given that male frogs aggregate and produce sounds to attract conspecific females, 
the duration and size of the choruses likely determine the mate attraction performance of the whole aggregation
(see Discussions for details). 
We quantify this feature by using {\it energy depletion time} and {\it chorus size}.
Specifically, {\it energy depletion time} is defined 
as the duration until the energies of all the frogs have become less than $1\%$ of $E_{\mathrm{max}}$,
and {\it chorus size} is defined as the maximum number of calling frogs in each chorus.
Then, we varied the aggregation size between $10$ and $20$ model frogs
and examined how the two factors depend on the number of the satellite males.
Note that the number of the satellite males could be different in each trial of the simulations
because we randomized the positions of the males. 
Figure \ref{fig:LargeAggre_Quality}B-D shows the {\it energy depletion time} and {\it chorus size}
in the aggregations of $10$ frogs, $15$ frogs and $20$ frogs, respectively. 
While the energy depletion time is much longer in the aggregations with satellite males 
than in the aggregations with no satellite males
(the top panels of Figure \ref{fig:LargeAggre_Quality}B-D), 
the chorus size divides into two levels depending on the number of satellite males
(the bottom panels of Figure \ref{fig:LargeAggre_Quality}B-D).

% INSERT FIGURE
%%

\clearpage

\section{Discussion}
\label{sec:discussion}

The roles of excitatory and inhibitory interactions in aggregations of male frogs were theoretically studied 
from the viewpoint of energy efficiency and chorus activity.
First, we proposed a hybrid dynamical model in which male frogs switch 
their behavioral state % among the calling state, resting state and satellite state 
based on their internal condition and the interactions with other males. 
In particular, the effect activating the calling behavior of other males 
was treated as an excitatory interaction 
while the effect inducing the satellite behavior of neighbors
was treated as an inhibitory interaction.
Second, we performed numerical simulations on the assumption of different aggregation sizes.
The simulation of a small aggregation (three frogs) reproduced
both the collective chorus and satellite behavior 
that are observed in the aggregations of actual male frogs.
The simulation of large aggregations ($10$-$20$ frogs) demonstrated that 
(1) the energy depletion time is prolonged due to the existence of satellite males
and (2) the chorus activity is divided into two levels over the whole chorusing period.
Consequently, this theoretical study indicates that 
satellite males can contribute to the energy efficiency of the frog aggregations 
by splitting the maximum chorus activity over the whole period.

In the proposed model, the parameter $\Delta N$ describes the threshold of the difference of call number
within which male frogs transit to the satellite state.
Numerical simulation of the small aggregation demonstrated that the occurrence of satellite behavior depends on this parameter
(Figure \ref{fig:SmallAggre_ParaDep}).
Namely, model frogs often transit to the satellite state when the parameter $\Delta N$ is small, 
but they seldom transit to the satellite state when $\Delta N$ is large. 
Combined with the fact that the features of satellite behavior vary among frog species \cite{Gerhardt_2002},
this parameter is essential to further understand the origin of the behavioral variance.
For instance, a small value of $\Delta N$ corresponds to species 
that are sensitive to the difference of call number with their competitor, 
resulting in frequent transitions into the satellite state; 
a large value of $\Delta N$ corresponds to species 
that are insensitive to the difference of call number,
resulting in rare transitions into the satellite state. 
Thus, this study using the mathematical model allows us to infer the mechanisms of  satellite behavior 
by comparing the numerical simulations with empirical studies.
However, playback experiments using loudspeakers demonstrated that 
other call traits such as loudness, frequency and call complexity 
also affect the preference of female frogs \cite{Ryan_1992}. 
Further extension of our model is necessary to more precisely formulate the mechanisms of satellite behavior 
although call number is one of the most dominant factors determining the attractiveness of the callers.

Numerical simulations of large aggregations indicate that 
satellite males can prolong the energy depletion time of the entire aggregation 
by splitting the chorus activity into two levels (Figure \ref{fig:LargeAggre_Quality}).
Because the calls of male frogs attract conspecific females that are spreading and moving around the breeding site,
the long chorus realized by the prolonged energy depletion time 
likely increases the total females attracted to the breeding site.
On the other hand, the devision of the chorus activity can be either an advantage or a disadvantage.
Playback experiments using loudspeakers demonstrated that various chorus sizes differently attract females 
\cite{Ryan_1981}.
If the two activity levels realized by the satellite males are both effective, 
the mate attraction performance is likely optimized. 
However, the effective chorus size can vary among species, 
and, therefore, further analyses combining numerical simulations and empirical studies are required 
to comprehensively evaluate the performance per aggregation and also per individual.

Our numerical simulation assumed a specific spatial structure, 
i.e. the linear distribution of male frogs along a water body 
(Figure \ref{fig:LargeAggre_concept}A). 
While this is a common distribution of male frogs as observed in {\it Hyla japonica} \cite{Aihara_2014, Aihara_2016}, 
{\it Rachophorus shlegelii} \cite{Aihara_2021, Bando_2016} and {\it Litoria chrolis} \cite{Aihara_2017},
more scattered patterns are possible \cite{Gerhardt_2002}.
To examine the robustness of our results, 
we performed additional simulations on the assumption of other distributions: 
(1) an almost linear distribution that was slightly scattered along a circular field 
and (2) a distribution that was scattered within a circular field. 
Both simulations demonstrated that the energy depletion time is prolonged in the aggregation with satellite males 
while the chorus activity is split into two levels 
(Figures S1 and S2 in Supplementary Information).
Thus, the results of the simulations hold in these distributions, 
indicating the robustness of the results.

Future directions of this study includes the application of the proposed model 
to engineering fields.
For instance, we previously claimed that the mathematical model of frog choruses 
can be applied to the autonomous distributed control of wireless sensor networks (WSNs) \cite{Aihara_2019}. 
WSNs are known as a sensing system in which 
many sensor nodes deliver a data packet to a specific node due to multi-hop communication \cite{Dressler_2007, Rawat_2014}.
In general, WSNs are constructed as follows: 
(1) many nodes are spatially distributed with non-uniform density to monitor a wide area,
(2) each node senses the surrounding condition and sends the information as a data packet to neighboring nodes, 
and (3) the nodes deliver the data packet to a sink node (a node that collects all the necessary information)
due to autonomous and multi-hop communication among neighboring nodes. 
Because each node is usually driven by a limited amount of battery, 
there are two factors essential to increase the performance of WSNs: 
{\it energy efficiency} and {\it reliable communication}.
We showed that the mathematical model reproducing the collective frog choruses
allows us to reduce energy consumption by collectively switching the active and inactive states of the sensor nodes,
and allows us to increase the reliability of communication by avoiding the collision of the data packets among neighboring nodes \cite{Aihara_2019}.
To further improve the energy efficiency of WSNs, we need to focus on the spatial distribution.
Namely, the deployed nodes can be locally dense because of their non-uniform distribution.
In such a locally dense area
only a single node needs to be active for the sensing and delivery of the data packet. 
Therefore, its neighboring nodes are redundant and can be powered off.
The inhibitory interaction of the proposed model induces the inactive state of close neighbors 
(Figures \ref{fig:Small_aggregation} and \ref{fig:LargeAggre_concept}),
which can further improve the energy efficiency of WSNs by powering off the redundant nodes. 
Note that a similar idea based on the satellite behavior of frogs was already proposed \cite{Mutazono_2012}. 
The novelty and advantage of our methodology is to establish the inactive state in a locally dense area 
as well as the synchronous switching between the active and inactive states over the whole network.

\section*{Acknowledgements}

We are grateful to T. Ishizuki for drawing a picture of satellite behavior (Figure \ref{fig:concept}C).

\section*{Author Contributions}

IA, DC, YH and MM designed the research; IA performed numerical simulations; 
IA, DC and MM wrote the paper.

\section*{Competing Interests}

We declare we have no competing interests.

\section*{Funding}

This study was supported by JSPS Grant-in-Aid for Young Scientists (No. 18K18005) 
and Grant-in-Aid for Scientific Research (B) (No. 20H04144).

%\bibliography{MS}

\clearpage

% Sample
%\begin{figure*}
%  \begin{center}
%	\includegraphics[width=1.0\textwidth]{Figure0.eps}
%	\end{center}
%	\caption{
%	
%	}
%	\label{fig:0}
%\end{figure*}

\end{document}